\documentstyle[pre,aps,preprint,epsfig]{revtex}
\tightenlines
\input{epsf.tex}
\begin{document}
\title{\bf Singular statistics}
\author{Eug\`ene Bogomolny$^1$, Ulrich Gerland$^{2}$ and 
        Charles Schmit$^1$}
\address{$^1$Laboratoire de Physique Th\'eorique et Mod\`eles
         Statistiques
	 \thanks{Unit\'e de Recherche de l'Universit\'e Paris XI et du CNRS 
	   (UMR 8626).}, 
  Universit\'e Paris-Sud, 91405 Orsay Cedex, France}
\address{$^{2}$Phys. Dep., UCSD,  La Jolla, CA 92093-0319, USA.}

\date{\today}

\maketitle


\begin{abstract}
We consider the statistical distribution of zeros of random meromorphic
functions whose  
poles are independent random variables. It is demonstrated that correlation functions 
of these zeros can be computed analytically and explicit calculations are performed 
for the 2-point correlation function. This problem naturally appears in e.g. 
rank-one perturbation of an integrable Hamiltonian and, in particular, when 
a $\delta$-function potential is added to an integrable billiard.   
\end{abstract}

\pagebreak

\section{Introduction}
The investigation of statistical properties of quantum energy levels of a given 
system is a long-standing problem (see e.g. \cite{porter}-\cite{bohigas}). 
According to the 
accepted conjectures energy levels of integrable systems behave as independent
random variables (i.e. they obey the Poisson statistics) \cite{berrytabor}
and those of generic chaotic systems follow the random matrix predictions
\cite{bohigasgiannoni}. The proof of these conjectures in the full generality 
is without doubt quite difficult and is still lacking though partial results
(concerning mostly integrable models) are available (see e.g. \cite{markloff}
and references therein).

But there are systems which are neither integrable nor completely chaotic
for which quantum energy levels are defined by an equation 
\begin{equation}
f(E)=0
\end{equation}
with a well defined (and simple) function $f(E)$.

In \cite{bogomolnybohigas} the case of polynomial equation 
\begin{equation}
f(E)=\sum_{n=0}^{N}a_n E^n
\end{equation}
has been considered and statistical properties of solution of 
$f(E)=0$ have been calculated provided $a_n$ be independent random variables.

The purpose of this paper is to consider the case of random meromorphic
functions of the form
\begin{equation}
f(E)=P(E)+\sum_{j=1}^{N}\frac{r_j}{E-e_j},
\label{f}
\end{equation}
where $P(E)$ is a polynomial and $e_j$, $r_j$ are, correspondingly, poles and 
residues of $f(E)$.

The natural example leading to the quantization condition in this form is the 
perturbation of a Hamiltonian by rank-one perturbation. If $H^{(0)}_{\mu \nu}$ is an 
unperturbed Hamiltonian then the Hamiltonian after perturbation is
\begin{equation}
H_{\mu \nu}=H_{\mu \nu}^{(0)}+v_{\mu}v_{\nu},
\end{equation}
where $v_{\mu}$ is a perturbation vector.

Solutions of the `Schr\"odinger' equation
\begin{equation}
H_{\mu \nu}\psi_{\nu}=E\psi_{\mu}
\end{equation}
can be expressed through solutions of unperturbed equation
\begin{equation}
H_{\mu \nu}^{(0)}\psi_{\nu}^{(0)}(n)=e_n\psi_{\mu}^{(0)}(n)
\end{equation}
as follows
\begin{equation}
\psi_{\mu}=\sum_{n}c_n\psi_{\mu}^{(0)}(n),
\end{equation}
where (up to a factor)
\begin{equation}
c_n=\frac{<v|\psi^{(0)}(n)>}{E-e_n}
\end{equation}
provided new eigenvalues, $E$, obey the following quantization condition
\begin{equation}
\sum_n\frac{|<v|\psi^{(0)}(n)>|^2}{E-e_n}=1.
\label{rangone}
\end{equation}
Here $<v|\psi^{(0)}(n)>=\sum_{\mu}v_{\mu}\psi_{\mu}^{(0)}(n)$.

This equation has the form of Eq.~(\ref{f}) with $P(E)$=const while unperturbed
energy levels play the role of poles, and the residues are projections of unperturbed 
wave functions in the directions of perturbation vector
\begin{equation}
r_n=|<v|\psi^{(0)}(n)>|^2.
\end{equation}
The addition of a $\delta$-function potential
\begin{equation}
V(x)=\lambda \delta (\vec{x}-\vec{x}_0)
\label{11}
\end{equation}
corresponds exactly to a rank-one perturbation. In this case (see e.g. 
\cite{albeverio}, \cite{seba}) Eq.~(\ref{rangone}) takes the form
\begin{equation}
\lambda \sum_n\frac{|\psi_n^{(0)}(\vec{x}_0)|^2}{E-e_n}=1,
\label{12}
\end{equation} 
where $\psi_n^{(0)}(\vec{x})$ and $e_n$ are eigenfunctions and eigenvalues of 
the problem without the $\delta$-function potential.

Another model which leads to similar equations is the Bohr-Mottelson model
\cite{bohrmottelson} which describes the interaction of one level (denoted
below by index 0) with all other levels. The model is defined by the
Hamiltonian
\begin{equation}
  H=H_0+V,
\label{new}  
\end{equation}  
where the interaction potential has non-zero matrix elements only between
the chosen level and all other levels
\begin{equation}
V_{0i}=V_{i0},\;\; V_{00}=V_{ij}=0.
\end{equation}
The energy levels of the Hamiltonian (\ref{new}) obey the equation \cite{bohrmottelson} 
\begin{equation}
\sum_{j}\frac{|V_{0j}|^2}{E-e_j}-(E-e_0)=0,
\end{equation}
which is again of the form of Eq.~(\ref{f}) with linear polynomial part.

A quite natural question appears: What is the statistical distribution of new
eigenvalues (i.e. solutions of Eq.~(\ref{f})) provided that statistical
distributions of poles and residues are known? In \cite{bogomolnyleboeuf}
it was proved that, if unperturbed system is described by random matrix
theory, the distribution of new eigenvalues will also be of random matrix type.

The main purpose of this paper is to compute analytically the statistical
distribution of solutions of Eq.~(\ref{f}) when the poles, $e_j$, are
independent random variables (i.e. they obey the Poisson statistics). We
shall show that in this case the resulting statistics exhibits the level
repulsion and differs from known  distributions.

The plan of the paper is the following. In Section \ref{general} the general
formalism is described. In Section \ref{meandensity} the calculation of the
mean density is presented. In Section \ref{2point} the 2-point correlation
function is computed when all residues, $r_j$, in Eq.~(\ref{f}) 
are the same. Generalization to different residues is disscused in Section
\ref{different}. As the exact expression of the 2-point correlation function
is cumbersome, in Section \ref{series} the series expansion of the results
is given. In Section \ref{limiting} the limiting behavior of the 2-point 
correlation function for small and large energy difference is obtained
without the knowledge of the exact solution. The details of the calculation
of a certain important integral are presented in Appendix.

\section{General formalism}\label{general}

We consider the most interesting case of Eq.~(\ref{f}) when the mean
separation of the poles is much smaller than a characteristic scale of
polynomial $P(E)$. Under such condition this polynomial can be considered as
a constant and after dividing by it Eq.~(\ref{f}) takes the form
\begin{equation}
   \sum_{j=1}^N\frac{r_j}{E-e_j}=1.
\label{main}   
\end{equation}
Our goal is to find the statistical distribution of solutions, $E$, of 
this equation provided $r_j$ are constants and $N$ numbers $e_j$ 
are independent random variables with a common 
distribution $d\mu(e)$  which for simplicity we choose as follows
\begin{equation}
   d\mu(e)=\left \{\begin{array}{ll}
   \frac{1}{2W}de &\mbox{if}\;-W\leq e_j\leq W,\\
   0 & \mbox{otherwise}\end{array}\right . .
\label{mu}   
\end{equation}
As the density of these poles is a constant they can be considered as
eigenvalues of a 2-dimensional integrable billiard and we shall often
call them  energy levels (or unperturbed energy levels). All our
calculations remain also valid  in more general case when the mean density
of poles is not a constant but is  not changed noticeable in the scale 
of the mean pole separation
(e.g for 3-dimensional integrable models). The only difference is that
$N/2W$ below should be substituted by the local mean density of poles,
$\bar{\rho}$ (see the end of Section \ref{meandensity}). 

In general, if one is interested in solutions of equation
\begin{equation}
f(x_n)=0,
\end{equation}
it is often convenient  to express the exact density of such solutions 
\begin{equation}
   \rho(x)=\sum_n\delta(x-x_n),
\end{equation}   
in the following manner
\begin{equation}
   \rho(x)=\delta (f(x))|\frac{df(x)}{dx}|.
\end{equation}
The main advantage of such representation is the possibility of calculating
the statistical distribution of roots, $x_n$, directly from statistical
distribution of coefficients of $f(x)$. E.g. this method has been used for
deriving the distribution of roots  of random polynomials 
\cite{bogomolnybohigas}.

In our case 
\begin{equation}
   \rho(E)=\delta(\sum_{j=1}^N\frac{r_j}{E-e_j}-1)
   \sum_{k=1}^N\frac{r_j}{(E-e_j)^2}.
\end{equation}
Representing the $\delta$-function as the Fourier integral (i.e. considering
the characteristic function of the roots) one gets
\begin{equation}
   \rho(E)=\int_{-\infty}^{\infty}\frac{d\alpha}{2\pi}
   \exp (i\alpha(\sum_{j=1}^N\frac{r_j}{E-e_j}-1))
   \sum_{k=1}^N\frac{r_k}{(E-e_k)^2}.   
\end{equation}
It is this representation of the exact density  that we shall use throughout
the paper.

As all $e_j$ are considered as independent random variables this expression
can be rewritten in the form 
\begin{equation}
   \rho(E)=\int\frac{d\alpha}{2\pi}e^{-i\alpha}
   \prod_{j=1}^N\exp (i\frac{\alpha r_j}{E-e_j})
   \sum_{k=1}^N\frac{r_k}{(E-e_k)^2},   
\end{equation}
where all factors are also independent random variables
which clearly permits to find all mean values by straightforward integration.

\section{Mean density}\label{meandensity}

Let us start with the calculation of the mean density 
\begin{equation}
   <\rho(E)>=\int\frac{d\alpha}{2\pi}e^{-i\alpha}\prod_{j=1}^N\int d\mu(e_j)
   \exp (i\frac{\alpha r_j}{E-e_j})
   \sum_{k=1}^N\frac{r_k}{(E-e_k)^2}.
\end{equation}
The integrals can be transformed in the following way
\begin{equation}
   <\rho(E)>=\int\frac{d\alpha}{2\pi}e^{-i\alpha}
   \left (\sum_{k=1}^Nr_kg(r_k\alpha))\prod_{j\neq k}f(r_j\alpha) \right ),
\end{equation}
where
\begin{equation}
   f(\alpha)=\int d\mu(e) \exp (i\frac{\alpha}{E-e}),
\end{equation}
and
\begin{equation}
   g(\alpha)=\int d\mu(e) \frac{1}{(E-e)^2}\exp (i\frac{\alpha}{E-e})=
   -\frac{\partial^2}{\partial \alpha^2}f(\alpha).   
\end{equation}
Let us rewrite the expression for $f(\alpha)$ in the form
\begin{equation}
   f(\alpha)=1-\frac{1}{2W}I(\alpha),
\end{equation}
where
\begin{equation}
  I(\alpha)=\int_{-W}^W de(1-\exp (i\frac{\alpha}{E-e})),
\label{29}  
\end{equation}
As
\begin{equation}
g(\alpha)=\frac{1}{2W}\frac{\partial^2}{\partial \alpha^2}I(\alpha),   
\end{equation}
it is necessary to compute only $I(\alpha)$.

Though the above steps are exact for finite $N$, the most interesting case
is the case $N\rightarrow \infty$. In this limit only small
values of $\alpha$ are important ($\alpha \approx 1/N$) 
and it is necessary to take into account in $I(\alpha)$ only terms linear in
$\alpha$.

Due to the singular character of the integral $I(\alpha)$ (\ref{29}) 
one cannot just expand the
integrand in power of $\alpha$. If $E$ belongs to the support of the
measure, $-W<E<W$, the change of variable 
\begin{equation}
t=\frac{1}{E-e},
\end{equation}
reduces the integral for $I(\alpha)$ (\ref{29}) to a sum of two integrals
\begin{equation}
  I(\alpha)=\left (\int_{(E+W)^{-1}}^{\infty}+\int^{(E-W)^{-1}}_{-\infty}
  \right )(1-e^{i\alpha t})\frac{dt}{t^2},
\end{equation}
which can be transform as follows
\begin{equation}
  I(\alpha)=\left (\int_{-\infty}^{\infty}-\int^{(E+W)^{-1}}_{(E-W)^{-1}}
  \right )(1-e^{i\alpha t})\frac{dt}{t^2}.
\end{equation}
The first integral equals $\pi|\alpha |$ and in the second integral one can
safely use  perturbation theory in $\alpha$. The final result is
\begin{equation}
  I(\alpha)=\pi|\alpha |+i\alpha \ln \frac{W-E}{W+E} 
            +\alpha^2\frac{W}{E^2-W^2}  +O(\alpha^3),
\end{equation}
and 
\begin{equation}
  g(\alpha)=\frac{\pi}{W}\delta(\alpha)+\frac{1}{E^2-W^2}.
\end{equation}
For small values of $\alpha$
\begin{equation}
  e^{-i\alpha}\prod_{j=1}^Nf(r_j\alpha)=
  \exp (-\frac{N}{2W}v(\pi|\alpha |+i\frac{\alpha}{v'})),
\end{equation}
where $v$ plays the role of a `bare' coupling constant,
\begin{equation}
v=\frac{1}{N}\sum_{j=1}^N r_j,
\label{vbare}
\end{equation}
and $v'$ is a `renormalized' coupling constant
\begin{equation}
  \frac{1}{v'}=\frac{2W}{Nv}+\ln \frac{W-E}{W+E}.
\label{vprime}  
\end{equation}
The necessity of  renormalization for such type of equations is well known
when a $\delta$-function potential is added to a $d$-dimensional system with
$d\geq 2$ (see e.g. \cite{albeverio} and Eqs.~(\ref{98} and (\ref{131}))
where it is connected with  one-parameter self-adjoint extension of a singular
Hamiltonian. Physically the renormalization means that the limit of infinite
small size impurity is not uniquely  defined and depends on internal
details of the scatterer. All physically measurable quantities (like the
cross-section) depend only on renormalized coupling constant, $v'$. The bare
coupling constant, $v$, is not observable and can be arranged to produce any
$v'$. When a specific model of small-size scatterer is considered (e.g. a
hard disk with a small radius) one gets a concrete form of the bare (and
renormalized) coupling constant. Below we consider the most interesting case
when renormalised coupling constant is assumed to be independent of $N$ (or
energy). All other limits can be derived from this one.
Note that in our calculations the appearance of such renormalization (i.e.
the fact that the bare coupling constant, $v$, and the renormalization factor 
$\log(W-E)/(W+E)$ appear only in the combination (\ref{vprime})) is automatic.

Finally, when $-W<E<W$ the density of state is the sum of two
terms
\begin{equation}
  \rho_{in}(E)=\frac{N}{2W}-\frac{2W}{(W^2-E^2)(\pi^2 +1/v'^2)}.
\label{rho}  
\end{equation}
As $N$ is assumed to be large, the first term dominates and the mean density of
levels is 
\begin{equation}
\bar{\rho}=\frac{N}{2W},
\end{equation}
as it should be.

When $E$ is beyond the interval $[-W,W]$ the calculation is simpler as in
this case there is no singularity on the contour of integration and one can
simply expand the integrand of $I(\alpha)$ on series of $\alpha$
\begin{equation}
  I(\alpha)=i\alpha \ln \frac{E-W}{E+W}+\alpha^2\frac{W}{E^2-W^2} +O(\alpha^3).  
\end{equation}
Therefore
\begin{equation}
  \rho_{out}(E)=\frac{|\phi(E)'|}{\sqrt{2\pi}\sigma}
  \exp(-\frac{\phi^2(E)}{2\sigma^2}),
\end{equation}
where
\begin{equation}
  \phi(E)=\ln \frac{E+W}{E-W}-\frac{1}{v'},\;
  \sigma^2=\frac{4W^2}{(E^2-W^2)N}.
\end{equation}
When $N\rightarrow \infty$, $\sigma \rightarrow 0$ and 
\begin{equation}
\rho(E)\rightarrow \delta (E-E_c),
\end{equation}
where $E_c$ is a root of equation $\phi(E_c)=0$
\begin{equation}
E_c=W\coth \frac{1}{2v'}.
\end{equation}
These results correspond exactly to what one attends from the simple geometrical
picture of roots of Eq.~(\ref{main}). The poles, $e_j$, divide the real axis
into $N+1$ intervals.  Due to the pole behavior each interval contains one of 
solutions, $E$. There is only one eigenvalue outside of the support of the initial
measure
and all other $(N-1)$ eigenvalues are distributed practically uniformly inside the
initial interval $[-W,W]$. The second term in Eq.~(\ref{rho}) is a smooth
bump which is necessary to insure that
$$\int_{-W}^{W}\rho_{in}(E)dE=N-1,$$
which can  easily be checked by noting that $2W/(E^2-W^2)=\partial
(1/v')/\partial E$. 

In Eq.~(\ref{mu}) we have assumed the particular form of the distribution of
$d\mu(e)$ but the results will be valid for any form of this measure 
(provided that it is not changed noticeably in the scale of the mean distance
between levels) with the
substitution $N/2W\rightarrow \bar{\rho}$,  $E+W\rightarrow E-E_{min}$ and
$W-E\rightarrow E_{max}-E$ where $\bar{\rho}$ is the local mean density of
unpertubated levels, $E_{min}$ and $E_{max}$ are minimal and maximal values
of levels included in the sum (\ref{main}).

\section{2-point correlation function}\label{2point}

Using the previously discussed method one can compute higher correlation 
functions as
well. We consider here the calculation of the 2-point correlation function,
$R_2(E_1,E_2)$, defined in the standard way
\begin{equation}
R_2(E_1,E_2)=<\rho(E_1)\rho(E_2)>,
\end{equation}
where $<\ldots >$ denotes the mean value over all random variables.

For clarity we consider first the case where all residues are equal, $r_j=v$.
This case appears e.g. when a $\delta$-function potential is added to a
rectangular billiard with periodic boundary conditions (see Eq.~(\ref{98})).
More general case with different $r_j$ will be considered shortly in Section
\ref{different}.

When all residues are the same our defining equation takes the form
\begin{equation}
\sum_{j=1}^N\frac{1}{E-e_j}=\frac{1}{v},
\label{mainf}
\end{equation}
and the 2-point correlation function can be expressed as follows
\begin{eqnarray}
R_2(E_1,E_2)&=&<\int \frac{d\alpha_1 d\alpha_2}{4\pi^2}\exp (i\sum_{j=1}^N
(\frac{\alpha_1}{E_1-e_j}+\frac{\alpha_2}{E_2-e_j}))
\nonumber\\
&\times& \sum_{k_1,k_2=1}^{N}
\frac{1}{(E_1-e_{k_1})^2(E_2-e_{k_2})^2}e^{-\frac{i}{v}(\alpha_1+\alpha_2)}>.
\end{eqnarray}
After a simple algebra this expression can be transformed to
\begin{eqnarray}
  &&R_2(E_1,E_2)=\int \frac{d\alpha_1
  d\alpha_2}{4\pi^2}[N(f(\alpha_1,\alpha_2))^{N-1}g(\alpha_1,\alpha_2) 
  \label{2function}\\
  &&+N(N-1)(f(\alpha_1,\alpha_2))^{N-2}
  \psi_1(\alpha_1,\alpha_2)\psi_2(\alpha_1,\alpha_2)]\exp
  (-\frac{i}{v}(\alpha_1+\alpha_2)),\nonumber
\end{eqnarray}
where
\begin{eqnarray}
  f(\alpha_1,\alpha_2)&=&\int d\mu(e)\exp(i\frac{\alpha_1}{E_1-e}+
  i\frac{\alpha_2}{E_2-e}),\nonumber\\ 
  g(\alpha_1,\alpha_2)&=&\int d\mu(e)\exp(i\frac{\alpha_1}{E_1-e}+
  i\frac{\alpha_2}{E_2-e})\frac{1}{(E_1-e)^2(E_2-e)^2},\nonumber\\
  \psi_1(\alpha_1,\alpha_2)&=&\int d\mu(e)\exp(i\frac{\alpha_1}{E_1-e}+
  i\frac{\alpha_2}{E_2-e})\frac{1}{(E_1-e)^2}
  \label{definition}, \\
  \psi_2(\alpha_1,\alpha_2)&=&\int d\mu(e)\exp(i\frac{\alpha_1}{E_1-e}+
  i\frac{\alpha_2}{E_2-e})\frac{1}{(E_2-e)^2}\nonumber.
\end{eqnarray}
We shall be interested in the distribution of eigenvalues inside the interval
$[-W,W]$ and therefore shall assume that both arguments $E_1$ and $E_2$
belong to this interval. 

Let us as  denote
\begin{equation}
  f(\alpha_1,\alpha_2)=1-\frac{1}{2W}I(\alpha_1,\alpha_2),
\end{equation}
where
\begin{equation}
  I(\alpha_1,\alpha_2)=\int_{-W}^{W}(1-\exp(i\frac{\alpha_1}{E_1-e}+
  i\frac{\alpha_2}{E_2-e}))de .
\label{mainI}  
\end{equation}
Other functions are expressed through $I(\alpha_1,\alpha_2)$ as follows
\begin{eqnarray}
  g(\alpha_1,\alpha_2)&=&-\frac{1}{2W}
  \frac{\partial^4}{\partial \alpha_1^2\partial
    \alpha_2^2}I(\alpha_1,\alpha_2),
  \nonumber\\
  \psi_1(\alpha_1,\alpha_2)&=&\frac{1}{2W}
  \frac{\partial^2}{\partial \alpha_1^2}I(\alpha_1,\alpha_2),\\
  \psi_2(\alpha_1,\alpha_2)&=&\frac{1}{2W}
  \frac{\partial^2}{\partial \alpha_2^2}I(\alpha_1,\alpha_2).\nonumber
\end{eqnarray}
The integral (\ref{mainI}) which defines $I(\alpha_1,\alpha_2)$ can be split
into three terms
\begin{equation}
I(\alpha_1,\alpha_2)=\left (\int_{-\infty}^{\infty}-\int_{-\infty}^{-W}-\int_{\infty}^{W}
\right )
(1-\exp(i\frac{\alpha_1}{E_1-e}+
  i\frac{\alpha_2}{E_2-e}))de .  
\end{equation}
In the first integral (which we denoted by $J(\alpha_1,\alpha_2)$)
singular points $E_1$ and $E_2$ are on the contour of integration. In the
second and the third ones  there are no singularities and they can be
computed in perturbation theory on $\alpha_1$ and $\alpha_2$. In the later
integrals we will see that one needs only linear in $\alpha$ terms and
\begin{equation}
  I(\alpha_1,\alpha_2)=J(\alpha_1,\alpha_2)+i(\alpha_1\ln \frac{W-E_1}{W+E_1}
  +\alpha_2\ln \frac{W-E_2}{W+E_2}).
\label{55}  
\end{equation}
It is the calculation of the first term which is difficult. The details of
this calculation are given in Appendix. The final result for 
$J(\alpha_1,\alpha_2)$ is the following
\begin{eqnarray}
 && J(\alpha_1,\alpha_2)=\pi(\alpha_1+\alpha_2)\mbox{sgn} (\alpha_2) -
  \pi [i(\alpha_1+\alpha_2)
  G(-\frac{\alpha_2}{\omega},\frac{\alpha_1}{\omega})
  \label{72}\\&&+ (\alpha_2  J_0(\xi)+i\sqrt{-\alpha_1\alpha_2}J_1(\xi))
\exp (i\frac{\alpha_1-\alpha_2}{\omega})]  
  (\mbox{sgn}(\alpha_1)-\mbox{sgn}(\alpha_2)),\nonumber
\end{eqnarray}
where $\omega=E_1-E_2$, $\xi=\frac{2}{\omega}\sqrt{-\alpha_1\alpha_2}$ and 
\begin{equation}
G(x,y)=e^{iy}\int_x^{\infty}J_0(2\sqrt{yt})e^{it}dt.
\end{equation}
The symmetry relations
\begin{eqnarray}
J(\alpha_2,\alpha_1)&=&J^*(\alpha_1,\alpha_2),\nonumber\\  
J(-\alpha_1,-\alpha_2)&=&J^*(\alpha_1,\alpha_2),\\  
J(-\alpha_2,-\alpha_1)&=&J(\alpha_1,\alpha_2),\nonumber
\end{eqnarray}
are also useful. 
We are interested in the situation when the difference of energies,
$\omega=E_1-E_2$, is of the order of the mean distance between the levels
\begin{equation}
\omega=\Omega\frac{2W}{N},
\end{equation}
and dimensionless frequency $\Omega$ is a constant. In this case one can
check that the important values of $\alpha$ will  also be of the order of $1/N$
which explains why we have restricted the expansion only up to linear terms.
Other simplification comes from the fact that in perturbation theory
terms (\ref{55}) one can put $E_1=E_2$ after which  they  only depend 
on the sum $\alpha_1+\alpha_2$.

In the limit of large $N$ one obtains (see Appendix)
\begin{eqnarray}
f^N(\alpha_1,\alpha_2)&=&\exp (-\frac{N}{2W}\tilde{I}(\alpha_1,\alpha_2)),\\
g(\alpha_1,\alpha_2)&=&\frac{1}{2W}\frac{1}{\omega^2}
(\frac{\partial}{\partial \alpha_1}-\frac
  {\partial}{\partial \alpha_2})
   [\exp(i\frac{\alpha_1-\alpha_2}{\omega})\Phi(\alpha_1,\alpha_2)],\\
\psi_1(\alpha_1,\alpha_2)&=&\frac{1}{2W}   
\exp(i\frac{\alpha_1-\alpha_2}{\omega})\frac {\partial}{\partial \alpha_1} 
   \Phi(\alpha_1,\alpha_2),\\
\psi_2(\alpha_1,\alpha_2)&=&-\frac{1}{2W} 
   \exp(i\frac{\alpha_1-\alpha_2}{\omega})\frac {\partial}{\partial \alpha_2} 
   \Phi(\alpha_1,\alpha_2).
\end{eqnarray}
Here we introduce 
\begin{equation}
  \tilde{I}(\alpha_1,\alpha_2)=J(\alpha_1,\alpha_2)
  +(\alpha_1+\alpha_2)\frac{i}{v'},
\end{equation}
where $v'$ is the renormalized coupling constant as in (\ref{vprime}) and 
\begin{equation}
\Phi(\alpha_1,\alpha_2)=2\pi  
J_0((\frac{2}{\omega}\sqrt{-\alpha_1\alpha_2})
\Theta(-\alpha_1\alpha_2)\mbox{sgn}(\alpha_1).
\end{equation}  

Therefore
\begin{eqnarray}
  R_2(\omega)&=&\int \frac{d\alpha_1d\alpha_2}{(4\pi W)^2}
  \{Nf^{N-1}\frac{2W}{\omega^2}
  (\frac{\partial}{\partial \alpha_1}-\frac{\partial}{\partial \alpha_2})
  [e^{i\frac{\alpha_1-\alpha_2}{\omega}}\Phi]
\nonumber\\
&-&
  N(N-1)f^{N-2}e^{2i\frac{\alpha_1-\alpha_2}{\omega}}
  [\frac {\partial}{\partial \alpha_1}\Phi]
  [\frac {\partial}{\partial\alpha_2}\Phi]\}.
\end{eqnarray}
It is convenient to  integrate the first term by parts 
\begin{eqnarray}
  &&\int f^{N-1}(\frac{\partial}{\partial \alpha_1}-\frac{\partial}{\partial \alpha_2})
  [e^{i\frac{\alpha_1-\alpha_2}{\omega}}\Phi]
=\nonumber\\
&&
\frac{N-1}{2W}\int f^{N-2}e^{i\frac{\alpha_1-\alpha_2}{\omega}}\Phi
(\frac{\partial}{\partial \alpha_1}-\frac{\partial}{\partial
  \alpha_2})J=
\\
&&\frac{N-1}{2W}\int f^{N-2}e^{2i\frac{\alpha_1-\alpha_2}{\omega}}\Phi^2.
\nonumber
\end{eqnarray}
Substituting this expression into the previous equation one obtains
\begin{equation}
R_2(\omega)=\frac{N(N-1)}{(4\pi W)^2}\int
d\alpha_1d\alpha_2\{\frac{\Phi^2}{\omega^2}-
[\frac {\partial}{\partial \alpha_1}\Phi]
[\frac {\partial}{\partial\alpha_2}\Phi]\} f^{N-2}
e^{2i\frac{\alpha_1-\alpha_2}{\omega}}.
\label{r2}
\end{equation}
The second useful form can be derived by the following transformation of 
the second term 
\begin{equation}
  e^{\Psi}[\frac {\partial}{\partial \alpha_2}\Phi]
  [\frac {\partial}{\partial\alpha_1}\Phi]=  
  [\frac{\partial^2}{2\partial \alpha_1\partial \alpha_2}
  +\frac{i}{\omega}
  (\frac{\partial}{\partial \alpha_1}-\frac{\partial}{\partial \alpha_2})
  +\frac{1}{\omega^2}]\Phi^2e^{\Psi},
\end{equation}  
where 
\begin{equation}
\Psi=2i\frac{\alpha_1-\alpha_2}{\omega}.
\end{equation}
Combining these two expressions one gets
\begin{equation}
R_2(\omega)=-\frac{N(N-1)}{(4\pi W)^2}\int d\alpha_1d\alpha_2 
e^{-\frac{N}{2W}\tilde{I}}[\frac{\partial^2}{2\partial \alpha_1\partial \alpha_2}
+\frac{i}{\omega}
(\frac{\partial}{\partial \alpha_1}-\frac{\partial}{\partial \alpha_2})]
\Phi^2e^{\Psi}.
\end{equation}
It is easy to check that under the scale transformation (assuming $\lambda >0$)
\begin{equation}
\omega\rightarrow \lambda \omega,\; \alpha_i\rightarrow \lambda \alpha_i,
\end{equation}
the pre-factor does not change and $\tilde{I}\rightarrow \lambda \tilde{I}$.
Therefore after the transformations
\begin{equation}
\Omega=\frac{N}{2W}\omega,
\end{equation}
and 
\begin{equation}
R_2(\omega)=\frac{N(N-1)}{4W^2}r_2(\Omega),
\end{equation}
plus the corresponding change of $\alpha$ the dependence of $N$ will
disappear and after the substitution
\begin{equation}
\alpha_i=\Omega\alpha_i
\end{equation}
the resulting expression for the 2-point correlation function  takes
the form
\begin{equation}
r_2(\Omega)=-\int \frac{d\alpha_1d\alpha_2}{4\pi^2} 
e^{-2\pi\Omega\tilde{J}}[\frac{\partial^2}{2\partial \alpha_1\partial \alpha_2}
+i
(\frac{\partial}{\partial \alpha_1}-\frac{\partial}{\partial \alpha_2})]
\Phi^2e^{2i(\alpha_1-\alpha_2)}.
\label{99}
\end{equation}  
where
\begin{equation}
\tilde{J}(\alpha_1,\alpha_2)=\bar{J}(\alpha_1,\alpha_2)
+i(\alpha_1+\alpha_2)\frac{1}{2\pi v'},
\label{jtilde}
\end{equation}
and from Eq.~(\ref{functional})
\begin{eqnarray}
& &\bar{J}(\alpha_1,\alpha_2)=
-\frac{1}{2}(\alpha_1+\alpha_2)\mbox{sgn}(\alpha_2)\nonumber\\
&+&\frac{i}{2}[(\alpha_1+\alpha_2)
  G(\alpha_1,-\alpha_2)
 - (i\alpha_1J_0(2\sqrt{-\alpha_1\alpha_2})\label{jbar}  \\
&+&\sqrt{-\alpha_1\alpha_2}J_1(2\sqrt{-\alpha_1\alpha_2}))
  e^{i(\alpha_1-\alpha_2)}]
  (\mbox{sgn}(\alpha_1)-\mbox{sgn}(\alpha_2)).
  \nonumber
\end{eqnarray}
When $\Omega\rightarrow 0$ it is convenient to perform the integration by
part 
\begin{equation}
  r_2(\Omega)=-\int d\alpha_1d\alpha_2 
  e^{-2\pi\Omega\tilde{J}}[\frac{\Omega^2}{2}
  \frac{\partial \tilde{J}}{\partial \alpha_1}
  \frac{\partial \tilde{J}}{\partial \alpha_1}
  +\frac{3\Omega i}{8\pi^2}\Phi e^{i(\alpha_1-\alpha_2)}]
   \Phi^2e^{2i(\alpha_1-\alpha_2)},
\end{equation}    
and take into account only the linear in $\Omega$ term
\begin{equation}
r_2(\Omega)\rightarrow \Omega A,
\label{smallomega}
\end{equation}
where 
\begin{equation}
A=-\frac{3 i}{8\pi^2}\int d\alpha_1d\alpha_2 \Phi^3e^{3i(\alpha_1-\alpha_2)},
\end{equation}
As in the region 
$\alpha_1\alpha_2<0$ $\Phi=2\pi J_0 (2\sqrt{-\alpha_1\alpha_2})$,
after the change of variables
\begin{equation}
\xi=2\sqrt{-\alpha_1\alpha_2},\;\;\eta=-\frac{\alpha_1}{\alpha_2},
\end{equation}
one gets
\begin{equation}
  A=-\frac{3\pi i}{2}\int_0^{\infty} \xi J_0^3(\xi)d\xi \int_0^{\infty}
  \frac{d\eta}{\eta} e^{\frac{3\xi i}{2}(\eta+\eta^{-1})} +c.c. .
\label{105}  
\end{equation}
The integral over $\eta$ equals $i\pi H_0^{(1)}(3\xi)$ (see (\ref{56}))
and the final expression for $A$ is
\begin{equation}
  A=3\pi^2 \lim_{\epsilon\rightarrow 0}
  \int_0^{\infty} \xi J_0((3+\epsilon)\xi) J_0^3(\xi)d\xi.
\end{equation}
We write here $(3+\epsilon)$ (where $\epsilon$ is proportional to $\Omega$) as 
this integral is a discontinuous integral and its value when 
$\epsilon=0$ is a half of the value for $\epsilon\rightarrow 0$. The last
value can be computed using the following integral (\cite{3}, p. 414)
\begin{equation}
  \int_0^{\infty}\prod_{n=1}^{4}J_0(a_n t)tdt=
  \frac{1}{\pi^2 \sqrt{a_1a_2a_3a_4}}
    \left \{ \begin{array}{ll}
     K(x),\;&\mbox{if}\; x<1\\
     \frac{1}{x}K(x),\;&\mbox{if}\; x>1
   \end{array}\right . ,
\label{107}   
\end{equation}
where $K(x)$ is the full elliptic integral of the second kind
\begin{equation}
x=\frac{\Delta}{\sqrt{a_1a_2a_3a_4}}
\end{equation}
and 
\begin{equation}
16\Delta^2=\prod_{n=1}^4(a_1+a_2+a_3+a_4-2a_n).
\end{equation}
If the left hand side is negative the above integral equals zero. 

In our case $\Delta\rightarrow 0$ and $K(0)=\pi/2$, therefore
\begin{equation}
  \lim_{\epsilon\rightarrow 0}\int_0^{\infty} \xi J_0((3+\epsilon)\xi)
  J_0^3(\xi)d\xi=\frac{1}{2\pi \sqrt{3}}.
\end{equation}  
Hence
\begin{equation}
A=\frac{\pi \sqrt{3}}{2}\approx 2.72\ldots .
\label{111}
\end{equation}
Note that the slope at the origin is independent on the coupling constant
and differs from the prediction of the Gaussian Orthogonal Ensembles of
random matrices ($ r_2(\Omega)\rightarrow (\pi^2/6)\Omega $ \cite{mehta}).

To find the asymptotics of the 2-point correlation function when
$\Omega\rightarrow \infty$ it is convenient to use Eq.~(\ref{r2}).
After rescaling of this expression one obtains (the constant term
comes from the $\delta$-function contribution of derivatives) 
\begin{eqnarray}
r_2(\Omega)&=& 1+\left \{\int_0^{\infty} d\alpha_1 \int^0_{-\infty}d\alpha_2 
[J_0^2(2\sqrt{-\alpha_1\alpha_2})+J_1^2(2\sqrt{-\alpha_1\alpha_2})]\right .
\nonumber\\
&\times&
\left .\exp(-2\pi \Omega \tilde{J}+2i(\alpha_1-\alpha_2))+c.c.\right \}.
\label{r2new}
\end{eqnarray}
When $\Omega\rightarrow \infty$ the dominant contribution comes from region
of small $\alpha$. Taking into account that when $\alpha\rightarrow 0$
\begin{equation}
\tilde{J}\rightarrow   \frac{1}{2}(\alpha_1-\alpha_2)+
i\frac{\alpha_1+\alpha_2}{2\pi v'},
\end{equation}
one concludes that the corresponding asymptotics of the 2-point
correlation function is 
\begin{equation}
r_2(\Omega)\rightarrow 1+\frac{2}{\Omega^2(\pi^2+1/v'^{2})}.
\label{asr2}
\end{equation}
Note the absence of oscillation on large $\Omega$ typical for standard
random matrix ensembles.

To check the above results we compute the statistical distribution of energy
levels of a rectangular billiard with a $\delta$-function potential inside
(sometimes called the Seba billiard \cite{seba}). 

For a rectangle of sides $a$ and $b$ solutions of the Schr\"odinger equation
\begin{equation}
(e_{\vec{n}}-\Delta )\psi_{\vec{n}}(\vec{x})=0
\end{equation}
in 2 dimensions with periodic boundary conditions have the form
\begin{equation}
\psi_{\vec{n}}(\vec{x})=\frac{1}{\sqrt{ab}}\exp (i\frac{2\pi}{a}n x
+i\frac{2\pi}{b}m y),
\end{equation}
and
\begin{equation}
e_{\vec{n}}=(\frac{2\pi}{a}n)^2+(\frac{2\pi}{b}m)^2
\end{equation}
for all (positive and negative) integers $n$ and $m$.

As $|\psi_{\vec{n}}(\vec{x})|^2=1/ab$ for all levels
Eq.~(\ref{12}) which determines energy levels after the introduction of a
$\delta$-function potential (\ref{11}) takes the form
\begin{equation}
v\sum_{\vec{n}}\frac{1}{E-e_{\vec{n}}}=1
\label{sum}
\end{equation}
with  $v=\lambda/ab$. 

Unperturbed eigenvalues have multiplicity 4 (for non-zero $m$, $n$) due to
the existence of positive and negative values of $m$, $n$. To remove this
degeneracy we consider in the above sum only positive integers and to have
the same mean density ($\bar{\rho}=ab/4\pi$) we divide all eigenvalues by 4
after which  eigenvalues included in the sum are
\begin{equation}
e_{\vec{n}}=(\frac{\pi}{a}n)^2+(\frac{\pi}{b}m)^2
\end{equation}
and $m,n>0$.

The sum (\ref{sum}) formally diverges and for computation we consider the
following renormalization
\begin{equation}
\frac{v'}{\bar{\rho}}(\sum_{\vec{n}}\frac{1}{E-e_{n}}-
  \bar{\rho}\int_{E_{min}}^{E_{max}}de \frac{1}{E-e})=1,
\label{98}
\end{equation}
where $E_{min}$ and $E_{max}$ are minimal and maximal values of energy
included in the sum. The subtracted  integral (considered as principal
value) equals $\log (E_{\max}-E)/(E-E_{min})$ and one obtains the same
relation between bare and renormalized coupling constants as before (cf.
(\ref{vprime}))
\begin{equation}
\frac{1}{v'}=\frac{1}{\bar{\rho}v}+\log \frac{E_{\max}-E}{E-E_{min}}.
\end{equation}
We take $v'=1$ and compute 100000 energy levels for such model.
In  Fig.~\ref{nsperiodic} the cumulative nearest-neighbor distribution
of these levels, $N(s)$, is presented. This quantity equals the integral
over the nearest-neighbor distribution
\begin{equation}
N(s)=\int_0^{\infty}p(s')ds'
\end{equation}
and it is better defined numerically than the usual nearest-neighbor
distribution. On the same figure two other curves are presented. The dashed
line corresponds to the Wigner surmise for the cumulative nearest-neighbor
distribution in the Gaussian Orthogonal Ensemble (GOE) of random matrices
\cite{mehta}
\begin{equation}
N_{GOE}(s)=1-e^{-\pi s^2/4}.
\end{equation}
The thin solid line represents the cumulative nearest-neighbor distribution
for the so-called semi-Poisson model \cite{bogomolnygerland}, \cite{gerland} 
which serves as a reference point in models with intermediate statistics
\begin{equation}
N_{s.P}(s)=1-(2s+1)e^{-2s}.
\end{equation}
It is clearly seen that the cumulative nearest-neighbor distribution for
the Seba billiard is quite far from GOE result and it is in between the
semi-Poisson curve and the GOE one. 

\begin{figure}[t]
 \epsfig{figure=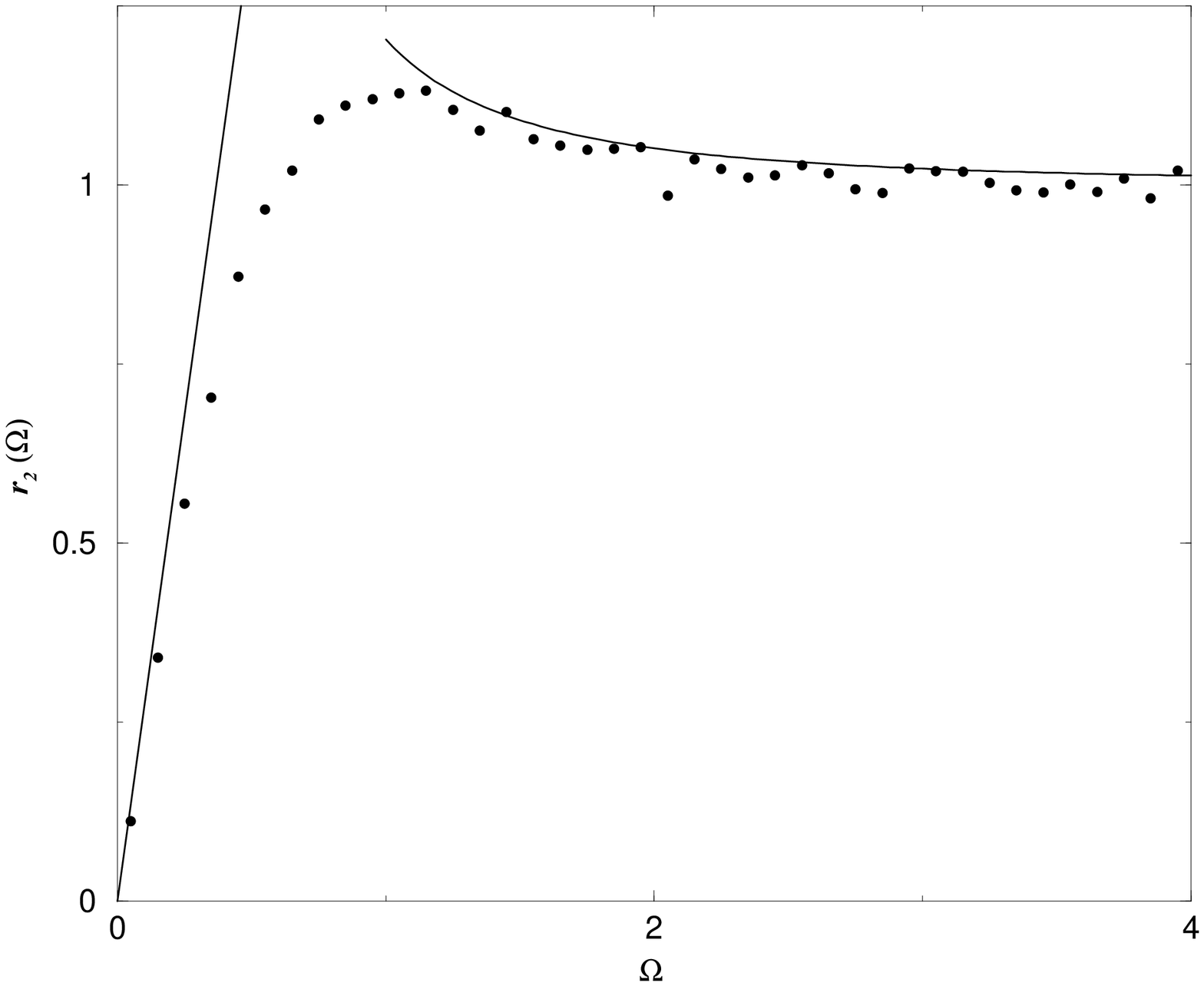,width=12cm,angle=270}
 \caption{The 2-point correlation function for the Seba billiard with the
   periodic boundary conditions.  Solid lines correspond to the asymptotics
   (\ref{smallomega}) and (\ref{asr2}) for small and large values of energy difference.}
   
 \label{r2periodic}
\end{figure}

The numerically computed 2-point correlation function for this model is
plotted in Fig.~\ref{r2periodic}. The two  curves in this figure 
correspond to theoretical
predictions for small and large values of the argument  given by
Eqs.~(\ref{smallomega}) and (\ref{asr2}) respectively. 

\section{Series expansions}\label{series}
The above expressions are quite cumbersome. Therefore it is of interest to
represent them as power expansions. We start with function $G(x,y)$
defined in (\ref{ge}). It is convenient to define 
\begin{equation}
G(x,y)=ig(t,s),
\end{equation}
where
\begin{equation}
t=ix, s=iy.
\end{equation}
Using the standard formula for the Bessel function
\begin{equation}
J_n(x)=\sum_{m=0}^{\infty}(\frac{x}{2})^{2m+n}\frac{1}{m!(m+n)!},
\end{equation}
one gets
\begin{eqnarray}
g(t,s)&=&\sum_{m=0}^{\infty}\frac{(-t)^m}{m!}
\sum_{n=m}^{\infty}\frac{(-s)^n}{n!}e^{t+s}
\nonumber \\
&=&1-\sum_{m=0}^{\infty}\frac{(-t)^m}{m!}
\sum_{n=0}^{m-1}\frac{(-s)^n}{n!}e^{t+s}.
\end{eqnarray}
Expanding the exponent leads
\begin{equation}
g(t,s)=1-\sum_{m,n=0}^{\infty}\frac{(-t)^m(-s)^n}{m!n!}R(m,n),
\end{equation} 
\begin{equation}
R(m,n)=\sum_{l=0}^nC_n^l(-1)^l\sum_{k=0}^{m-n+l-1}C_m^k(-1)^k,
\end{equation}
and $C_m^n$ are the binomial coefficients. Only terms for which the upper
limits in these sums are non-negative are included in the summation.
But
\begin{equation}
\sum_{k=0}^L C_m^k(-1)^k=(-1)^LC_{m-1}^L,
\end{equation}
therefore
\begin{equation}
R(m,n)=(-1)^{m+n-1}\sum_{l=0}^nC_n^lC_{m-1}^{m-n+l-1}=
(-1)^{m+n-1}C_{n+m-1}^n.
\end{equation}
Finally we get
\begin{equation}
g(t,s)=1+\sum_{m,n=0}^{\infty}\frac{t^ms^n}{m!n!}C_{n+m-1}^n.
\end{equation}
Using Eq.(\ref{jnew}) one can show that
\begin{equation}
-\frac{1}{\pi i \omega}J(\alpha_1,\alpha_2)=s+t
+2\sum_{n,m\geq 1}^{\infty}\frac{t^ms^n}{m!n!}C_{n+m-2}^{n-1},
\end{equation}
where, as before, $t=-i\alpha_2/\omega$ and $s=i\alpha_1/\omega$. 

The expansion of the pre-exponent factor in (\ref{r2}) can be simplified by
the following identity (\cite{3}, p.32)
\begin{equation}
J_n^2(z)=\sum_{m=0}^{\infty}\frac{(-1)^m(2m+2n)!}{m!(m+2n)![(m+n)!]^2}
(\frac{z}{2})^{2m+2n}.
\end{equation}
One gets
\begin{equation}
J_0^2(\xi)+J_1^2(\xi)=1+\sum_{m=1}^{\infty}
\frac{(2m)!}{[m!]^3(m+1)!}(\alpha_1\alpha_2)^m.
\end{equation}
Changing $\alpha_2\rightarrow -\alpha_2$ we can rewrite Eq.(\ref{r2new}) in
the form
\begin{eqnarray}  
r_2(\Omega)&=&1+\int_0^{\infty}d\alpha_1d\alpha_2 P(\alpha_1,\alpha_2)
\exp(-\Omega(\pi+\frac{i}{v'})\alpha_1-\Omega(\pi-\frac{i}{v'})\alpha_2)
\nonumber\\
& &\exp( 2i(\alpha_1+\alpha_2)+2\pi i\Omega Q(\alpha_1,\alpha_2))+c.c.,
\end{eqnarray}
where 
\begin{equation}
P(\alpha_1,\alpha_2)=1+\sum_{m=1}^{\infty}
\frac{(2m)!}{[m!]^3(m+1)!}(-\alpha_1\alpha_2)^m,
\end{equation}  
and
\begin{equation}
Q(\alpha_1,\alpha_2)=\sum_{n,m\geq 1}^{\infty}
i^{m+n}\frac{\alpha_2^m\alpha_1^n}{m!n!}C_{n+m-2}^{n-1}.
\end{equation}

\section{Limiting behavior}\label{limiting}

The above formulas give the exact expressions for the 2-point correlation
function for the problem considered but they are  quite cumbersome and suitable
mostly for numerical calculations. The most interesting information which
one can extract from them is the behavior of the 2-point correlation
function at small and large $\Omega$. The purpose of this section is to
discuss methods which permit to find these asymptotics without knowledge of 
the exact solution.  

It is clear that in order to find the behavior of the 2-point correlation
function in the limit $\omega\rightarrow 0$ it is necessary to consider only
the case when three initial levels (which we shall denote $e_1$, $e_2$,
$e_3$) are close each to others and all
other levels are far from this triple. In other words only three
terms in Eq.~(\ref{mainf}) are big. In such a case Eq.~(\ref{mainf})
which should determine the positions of two nearest levels can be approximated
as follows
\begin{equation}
\frac{1}{E-e_1}+\frac{1}{E-e_2}+\frac{1}{E-e_3}=0.
\end{equation}
(Note the absence of the coupling constant.) The solution of this equation is
\begin{equation}
E_{1,2}=\frac{e_1+e_2+e_3}{3}\pm\frac{1}{3}
\sqrt{e_1^2+e_2^2+e_3^2-e_1e_2-e_1e_3-e_2e_3}.
\end{equation}
This expression is translationally invariant, therefore one can assume that
$e_1+e_2+e_3=0$ and the difference between two adjacent levels is
\begin{equation}
\Delta E=\frac{2}{\sqrt{3}}
\sqrt{e_1^2+e_2^2-e_1e_2}.
\end{equation}
After the corresponding rescaling the 2-point correlation function at the
limit $\Omega\rightarrow 0$ takes
the form
\begin{equation}
r_2(\Omega)=\frac{1}{2}\int \delta(\Omega-\frac{2}{\sqrt{3}}
\sqrt{e_1^2+e_2^2-e_1e_2})de_1de_2.
\end{equation}
The factor $1/2$ comes from the restriction $e_1<e_2$. Changing variables
$e_1=r\cos \theta$, $e_2=r\sin \theta$ and performing the integral over $r$
one gets
\begin{equation}
r_2(\Omega)=\frac{3}{8}\Omega\int \frac{d\theta}
{1-\sin\theta \cos\theta}.
\end{equation}
The last integral equals $4\pi/\sqrt{3}$ and finally we obtain that in the
limit of small $\Omega$ 
\begin{equation}
r_2(\Omega)=\frac{\pi \sqrt{3}}{2}\Omega,
\end{equation}
which coincides with the result (\ref{111}) obtained above by different method.

To compute the behavior of the 2-point correlation function at large
$\Omega$ it is convenient to use a method based on the  usual trace formula

Let us define 
\begin{equation}
G(z)=\sum_{i=1}^{N}\frac{1}{z-e_i},
\end{equation}
where all $e_i$ are independent random variables as before.

We need to calculate the density of levels $E_j$ defined by the equation
\begin{equation}
v G(E_j)=1.
\end{equation}
Formally this density can be expressed in the following way
\begin{equation}
\rho(E)=-\frac{1}{\pi}\mbox{Im}[G(E)+\frac{\partial}{\partial E}\log
(1-v G(E))],
\label{trace}
\end{equation}
where the symbol Im$[F(E)]$ means the following limit
\begin{equation}
\mbox{Im}[F(E)] =\lim_{\epsilon\rightarrow 0}
\frac{1}{2i}(F(E+i\epsilon)-F(E-i\epsilon)),
\end{equation}
taken over positive $\epsilon$.

The derivation of Eq.(\ref{trace}) is simple. The function
$1-\lambda G(E)$ has zeros at $E_j$ and poles at $e_k$, therefore
\begin{equation}
\frac{\partial}{\partial E}\ln (1-v G(E))=\sum_j\frac{1}{E-E_j}-\sum_k
\frac{1}{E-e_k}.
\end{equation}
The first term in (\ref{trace}) cancels the poles from unperturbed levels and
the imaginary part produces $\delta$-function singularity at the required positions.

Let us denote 
\begin{equation}
G_{\pm}(E)=G(E\pm i\epsilon),
\end{equation}
where $\epsilon$ is positive and $\epsilon  \rightarrow 0$.

In calculating the mean values it is useful to take  explicitly into account
the mean values of $G_{\pm}(E)$. Using the relation 
\begin{equation}
\frac{1}{x\pm i\epsilon}=P\frac{1}{x}\mp i\pi \delta (x),
\end{equation}
one finds
\begin{equation}
<G_{\pm}(E)>=\mp \pi i \bar{\rho} +\bar{\rho}\log (\frac{W+E}{W-E}),
\label{131}
\end{equation}
where $\bar{\rho}$ is the mean level density of non-perturbed states.
Introducing
\begin{equation}
g_{\pm}(E)=G_{\pm}(E)-<G_{\pm}(E)>,
\end{equation}
one can write 
\begin{equation}
1-v G_{\pm}(E)=(1-v<G_{\pm}(E)>)(1-\lambda g_{\pm}(E)),
\end{equation}
where
\begin{equation}
\lambda=\frac{v'}{\bar{\rho}(1\pm i\pi v')},
\end{equation}
and $v'$ is the renormalized coupling constant the same as in
Eq.(\ref{vprime})
\begin{equation}
  \frac{1}{v'}=\frac{1}{\bar{\rho}v}+\ln \frac{W-E}{W+E}. 
\end{equation}
The density of states (ignoring small correction to the mean density of
state as in Eq.~(\ref{rho}) now will take the form
\begin{equation}
\rho(E)= \bar{\rho} -\frac{1}{2\pi i}(g_{+}(E)-g_{-}(E))-
\frac{1}{2\pi i}\frac{\partial}{\partial E}(\log (1-\lambda g_{+}(E))-
\log (1-\lambda^*g_{-}(E))).
\end{equation}
The 2-point correlation function is the mean value of the product of two such
expressions at different energies. The computation of the mean value can be
done in perturbation theory by expanding this expression into powers of $g_{\pm}(E)$
and using a formula
\begin{eqnarray}
<g_{+}^n(E_1)g_{-}^m(E_2)>&\approx& \bar{\rho}\int
\frac{de}{(E_1-e+i\epsilon)^n(E_2-e-i\epsilon)^m}\nonumber\\
&=&2\pi i(-1)^{n-1} \bar{\rho}
C_{m+n-2}^{n-1}\frac{1}{(\omega)^{m+n-1}}(1+O(\omega)),
\end{eqnarray}
where $\omega=E_1-E_2$. Therefore one can organize the perturbation series
in a series of inverse power of $\omega$. 

Taking into account the first terms in the expansion of the logarithm in the
above expression one gets
\begin{equation}
\rho(E)=\bar{\rho}-\frac{1}{2\pi i}[ (1-\lambda\frac{\partial}{\partial
  E})g_{+}(E)-(1-\lambda^*\frac{\partial}{\partial E})g_{-}(E)].
\end{equation}
At large $\omega$
\begin{eqnarray}
R_2(\omega)&=&\bar{\rho}^2 -\frac{1}{4\pi^2}
(\lambda \frac{\partial}{\partial E_1}g_{+}(E_1)g_{-}(E_2)
+(E_1\leftrightarrow E_2)+c.c.)\nonumber\\
&=& \bar{\rho}^2+\frac{2}{\omega^2(\pi^2+1/v'^2)},
\end{eqnarray}
which agrees with Eq.~(\ref{asr2}) derived from the general formula.

We stress that the methods used in this Section are not restricted to
particular cases considered. They also can be used in more general
situations where exact solution are not available, e.g. for rank-two
perturbations (2 short-range impurities) and similar problems. 

\section{General case}\label{different}
 
In the previous Sections we considered the calculation of the 2-point
correlation function under the assumption of equality of all residues. Here
the generalization of these calculations to the case of different residues
is presented. 

When the residues are different one has instead of Eq.~(\ref{mainf}) the
following equation
\begin{equation}
\sum_{j=1}^N\frac{r_j}{E-e_j}=1,
\end{equation}
and, consequently, instead of Eq.~(\ref{2function}) one gets a more general 
relation
\begin{eqnarray}
&&R_2(E_1,E_2)=\int_{-\infty}^{\infty} \frac{d\alpha_1 d\alpha_2}{4\pi^2}
e^{-i(\alpha_1+\alpha_2)}\sum_{k=1}^N r_k^2 g(r_k\alpha_1,r_k\alpha_2)
\prod_{j\neq k}f(r_j\alpha_1,r_j\alpha_2)\nonumber\\
&&+\sum_{k_1\neq k_2}r_{k_1}r_{k_2}\psi_1(r_{k_1}\alpha_1,r_{k_1}\alpha_2)
\psi_2(r_{k_2}\alpha_1,r_{k_2}\alpha_2)
\prod_{j\neq k_1,k_2}f(r_j\alpha_1,r_j\alpha_2),
\end{eqnarray}
where $f(\alpha_1,\alpha_2)$, $g(\alpha_1,\alpha_2)$, and
$\psi_i(\alpha_1,\alpha_2)$ are the same as in Eqs.~(\ref{definition}).
Repeating the same steps as in Section \ref{2point} one gets
exact expressions for the 2-point correlation function. The analog of
Eq.~(\ref{99}) (which is convenient for calculation of the small-$\Omega$
series of the 2-point correlation function) has the following form
\begin{eqnarray}
&&r_2(\Omega)=-\int \frac{d\alpha_1d\alpha_2}{4\pi^2} 
e^{-2\pi\Omega <\bar{J}(r\alpha_1,r\alpha_2)>-
  i\Omega v(\alpha_1+\alpha_2)/v'}\nonumber \\
&&\times
\left [\frac{\partial^2}{2\partial \alpha_1\partial \alpha_2}
<\Phi(r\alpha_1,r\alpha_2) e^{ir(\alpha_1-\alpha_2)}>^2 \right .\nonumber\\
&&+i(\frac{\partial}{\partial \alpha_1}-\frac{\partial}{\partial \alpha_2})
<\Phi(r\alpha_1,r\alpha_2) e^{ir(\alpha_1-\alpha_2)}>\nonumber \\
&&\times
\left .<\Phi(r\alpha_1,r\alpha_2) r e^{ir(\alpha_1-\alpha_2)}>\right ].
\label{new99}
\end{eqnarray}
Instead of Eq.~(\ref{r2new}) useful for large-$\Omega$ asymptotics one obtains
\begin{eqnarray}
&&r_2(\Omega)= 1+\{\int_0^{\infty} d\alpha_1 \int^0_{-\infty}d\alpha_2
e^{-2\pi \Omega <\bar{J}(r\alpha_1,r\alpha_2)>-i\Omega v(\alpha_1+\alpha_2)/v'} 
\label{newr2new}\\  
&&\times [<rJ_0^2(2r\sqrt{-\alpha_1\alpha_2})e^{ir(\alpha_1-\alpha_2)}>^2
\nonumber\\
&&+<rJ_1^2(2r\sqrt{-\alpha_1\alpha_2})e^{ir(\alpha_1-\alpha_2)}>^2]+c.c.\}
\nonumber 
\end{eqnarray}
Here $<f(r)>$ denotes the mean value over all residues
\begin{equation}
<f(r)>=\frac{1}{N}\sum_{j=1}^Nf(r_j),
\label{161}
\end{equation}
functions $\bar{J}(\alpha_1,\alpha_2)$ and $\Phi(\alpha_1,\alpha_2)$ are defined
in Eqs.~(\ref{jbar}) and (\ref{59}), $v$, and $v'$ are `bare' and `renormalized'
coupling constants (see (\ref{vbare}) and (\ref{vprime})). 

As in Section \ref{2point} it is of interest to compute the behavior of the
2-point correlation function at small and large energy difference. When
$\Omega \rightarrow 0$ the integration by parts as in Section \ref{2point}
leads to 
\begin{equation}
r_2(\Omega)\rightarrow \Omega A,
\label{162}
\end{equation}
where
\begin{eqnarray}
&&A=\int \frac{d\alpha_1 d\alpha_2}{4\pi^2}[\frac{i}{2}
<\Phi (r\alpha_1,r\alpha_2)r^2e^{ir(\alpha_1-\alpha_2)}>
<\Phi (r\alpha_1,r\alpha_2)e^{ir(\alpha_1-\alpha_2)}>^2\nonumber \\
&&+<\Phi (r\alpha_1,r\alpha_2)re^{ir(\alpha_1-\alpha_2)}>^2
<\Phi (r\alpha_1,r\alpha_2)e^{ir(\alpha_1-\alpha_2)}>.
\end{eqnarray}
Using (\ref{59}) this triple sum is transformed to the form (cf. with
Eq.~(\ref{105}))
\begin{eqnarray}
&&A=-i\frac{\pi}{N^3} \sum_{i,j,k=1}^N(\frac{1}{2}r_j^2+r_jr_k)
\int_0^{\infty}\xi d\xi J_0(r_i\xi)J_0(r_j\xi)J_0(r_k\xi)\nonumber\\
&&\times
\int_0^{\infty}\frac{d\eta}{\eta}e^{i(r_i+r_j+r_k)(\eta+1/\eta)\xi/2}+c.c
\end{eqnarray}
According to (\ref{56}) the last integral equals 
$i\pi H_0^{(1)}((r_i+r_j+r_k)\xi)$ therefore
\begin{equation}
A=\frac{2\pi^2}{N^3} \sum_{i,j,k=1}^N(\frac{1}{2}r_j^2+r_jr_k)
\int_0^{\infty}\xi d\xi J_0(r_i\xi)J_0(r_j\xi)J_0(r_k\xi)J_0((r_i+r_j+r_k)\xi).
\end{equation}
Taking into account Eq.~(\ref{107}) and symmetrising the answer one gets
\begin{equation}
A=\frac{\pi}{6}\frac{1}{N^3}
\sum_{i,j,k=1}^N\sqrt{\frac{(r_i+r_j+r_k)^3}{r_ir_jr_k}}.
\label{a}
\end{equation}
Of course, when all residues are equal, $A=\pi \sqrt{3}/2$ as in (\ref{111}).

When $\Omega\rightarrow \infty$ from Eq.~(\ref{newr2new}) one gets
\begin{eqnarray}
&& r_2(\Omega)\rightarrow
1+<r>^2\int_0^{\infty}d\alpha_1\int_{-\infty}^0d\alpha_2
e^{-\Omega <r>(\pi(|\alpha_1|+|\alpha_2|)-\frac{i}{v'}(\alpha_1+\alpha_2)}+c.c.
\nonumber\\
&&=1+\frac{2}{\Omega^2 (\pi^2+1/v'^2)},
\label{asr2d}
\end{eqnarray}
which differs from Eq.~(\ref{asr2}) only by a suitable definition of the
coupling constant.

Note that Eq.~(\ref{a}) is valid only for non-zero values of the residues.
Otherwise, the pre-factor $A$ formally diverges. This divergence is a consequence
of a simple fact that when some $r_j=0$ there exist certain energy levels
exactly equal unperturbed levels. Therefore the set of new energy
levels consists of two parts. The first includes energy levels which are
changed by the perturbation. Their correlation function is given by the
formulas above where only non-zero residues are taken into account. The
second part consists of energy levels which are not changed by the
perturbation. Their correlation functions are the same as for the Poisson
process and, in particular, they do not display level repulsion. As the
cross-correlations between (a finite number) of old and new energy levels
disappear when $N\rightarrow \infty$, the resulting statistics is a
superposition of two independent distributions and, in general, it will not
have level repulsion (i.e. $R_2(\epsilon) \neq 0$ when 
$\epsilon \rightarrow 0$).

The above case is realized e.g. when a $\delta$-function potential
(\ref{11}) is added to a rectangular billiard with the Dirichlet boundary
conditions and the positions of the singular point ($\vec{x}_0=(x_0,y_0)$)
are commensurable with the corresponding  sides of the rectangular ($a$ and $b$). 
In this model unperturbed wave functions are determined by two integers, $n$ and $m$,
\begin{equation}
\psi_{\vec{n}}=\frac{2}{\sqrt{ab}}\sin (\frac{\pi}{a}nx)\sin (\frac{\pi}{b}my)  
\end{equation}
and the residues are
\begin{equation}
r_{\vec{n}}=\frac{4}{ab}\sin^2 (\frac{\pi}{a}nx_0)\sin^2 (\frac{\pi}{b}my_0).   
\label{rn}
\end{equation}
If
\begin{equation}
\frac{x_0}{a}=\frac{p_1}{q_1},\;\;\frac{y_0}{b}=\frac{p_2}{q_2}
\label{ratio}
\end{equation}
for co-prime integers $p_i$ and $q_i$ there exist only a finite number of
different residues depending on values $n$ mod $q_1$ and $m$ mod $q_2$. In
particular, when $n$ is divisible by $q_1$ or $m$ is divisible by $q_2$,
$r_{\vec{n}}=0$. It means that for all these values of $n$ and $m$ wave
functions and energy eigenvalues will not be changed by the perturbation and
the resulting distribution (included all energy levels) will not
describe level repulsion. 

Another interesting case corresponds to a model when all residues are also
independent random variables with a probability $d\mu(r)$. If $r_j$ never
take very small values (more precisely, the mean value of $1/\sqrt{r}$ is
finite) the only
modification is that mean value over residues, $<f(r)>$, should be taken over
the given distribution i.e. instead of Eq.~(\ref{161}) one has to use
\begin{equation}
<f(r)>=\int f(r) d\mu(r).
\end{equation}
In particular the value of the pre-factor $A$ is
\begin{equation}
A=\frac{\pi}{6}\int d\mu(r_1) d\mu(r_2) d\mu(r_3)
\sqrt{\frac{(r_1+r_2+r_3)^3}{r_1r_2r_3}}.
\label{meana}
\end{equation}
But, if the probability of small values of residues
is large, certain expansions should be modified. A natural example is  e.g.
the Seba billiard with the Dirichlet boundary conditions when ratios of the
positions of the singularity to the corresponding sides (as in (\ref{ratio}))
are non-commensurable irrational numbers. In this case $r_n$ defined in 
(\ref{rn}) are equivalent to random variables 
\begin{equation}
r_{\vec{\phi}}=\frac{4}{ab}\sin^2 \phi_1\sin^2 \phi_2,
\label{rrandom}
\end{equation}
where angles $\phi_1$ and $\phi_2$ are  uniformly distributed between 0 and
$\pi$. 

\begin{figure}[t]
 \epsfig{figure=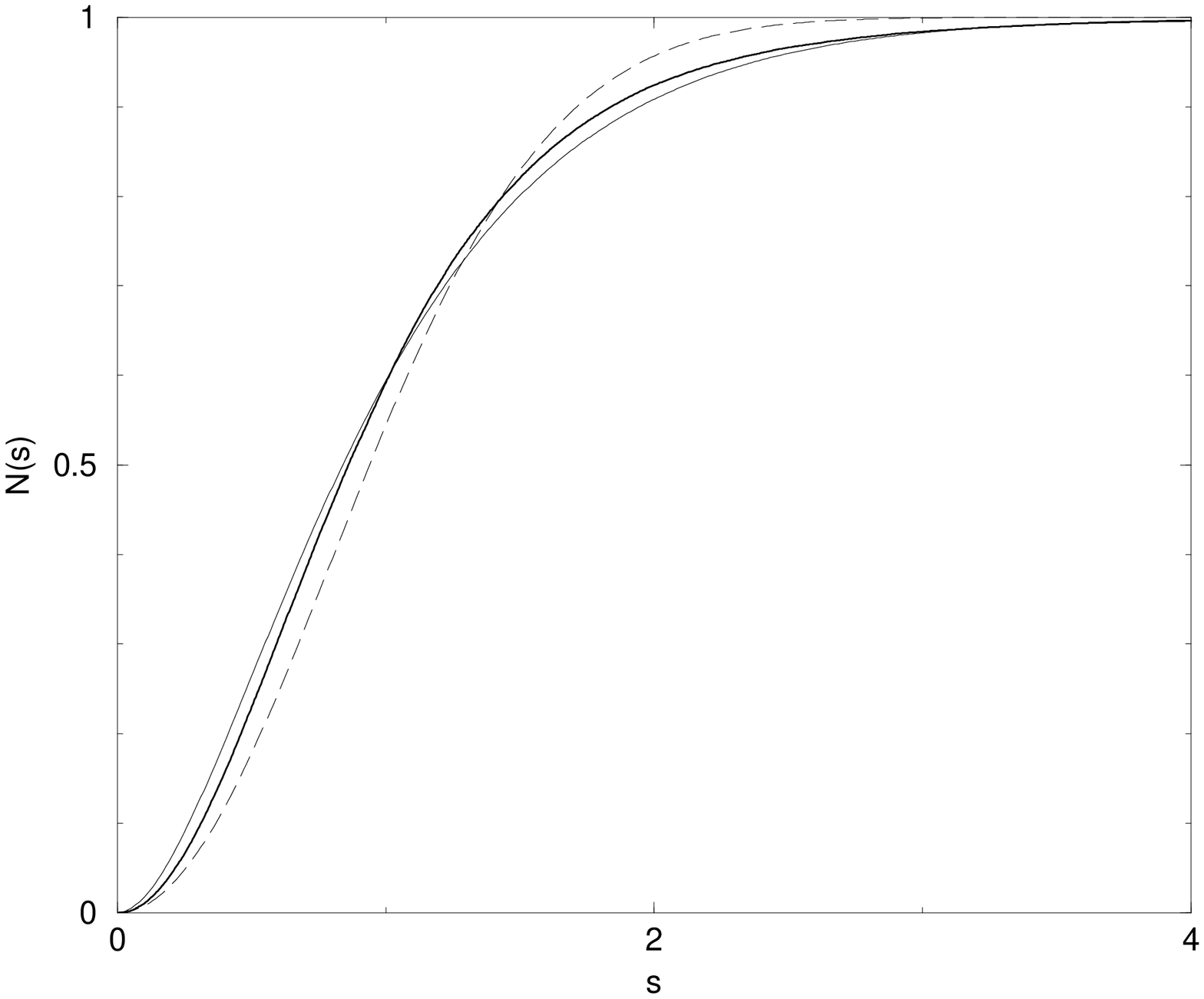,width=12cm,angle=270}
 \caption{Nearest-neighbor distribution for the Seba billiard with the periodic
   boundary conditions. Dashed line is the GOE result. Thin line is the
   semi-Poisson curve.}
 \label{nsperiodic}
\end{figure}

\begin{figure}[t]
 \epsfig{figure=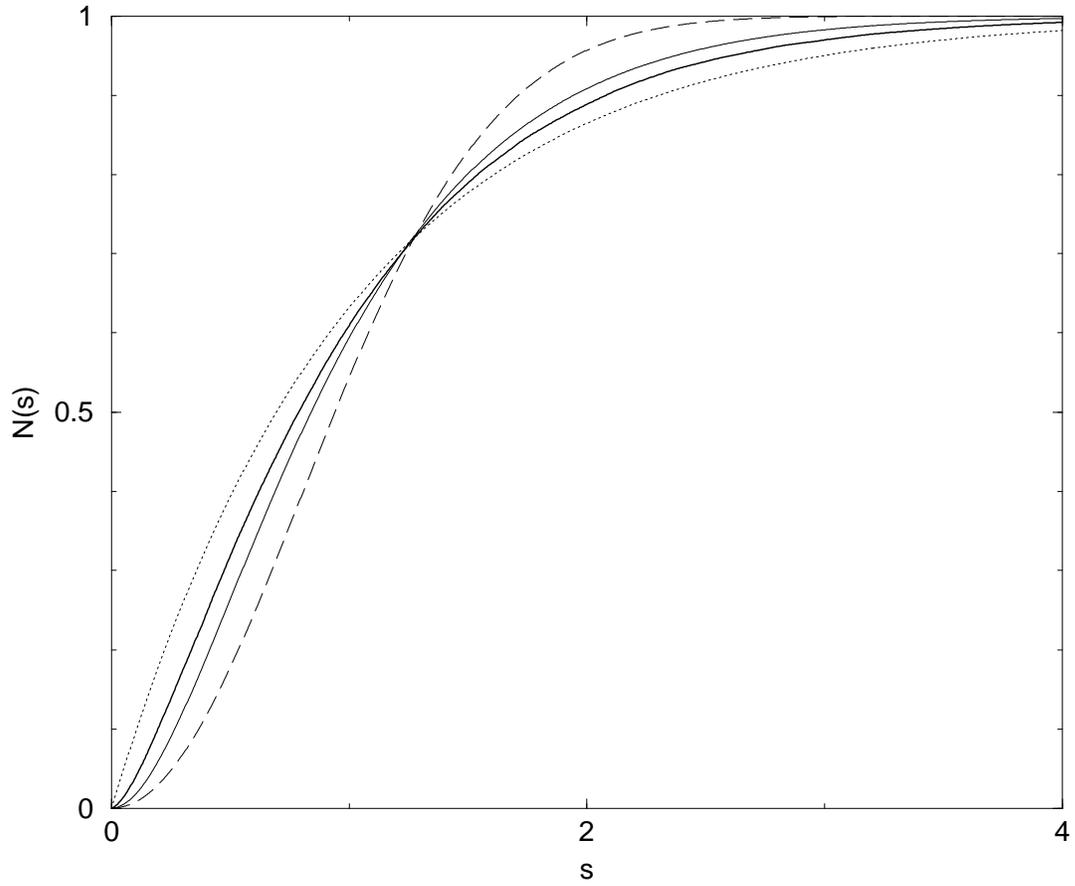,width=12cm,angle=270}
 \caption{Nearest-neighbor distribution for the Seba billiard with the Dirichlet
   boundary conditions. Dashed line is the GOE result. Thin line is the
   semi-Poisson curve. Dotted line is the Poisson prediction.}
 \label{nsdirichlet}
\end{figure}

\begin{figure}[t]
 \epsfig{figure=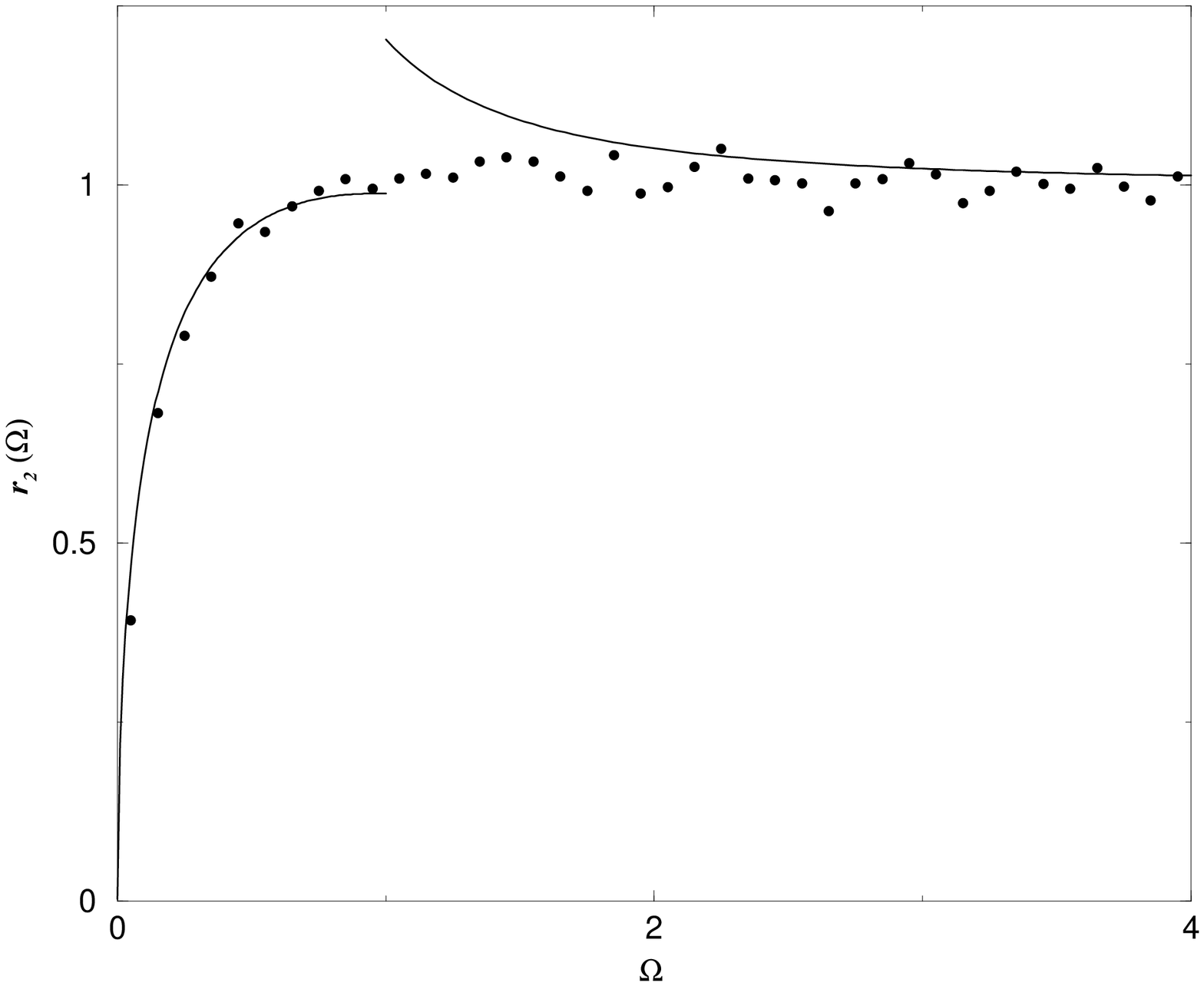,width=12cm,angle=270}
 \caption{The 2-point correlation function for the Seba billiard with
   the Dirichlet boundary conditions.  Solid lines correspond to the asymptotics
   (\ref{smallomegad}) and (\ref{asr2d}) for small and large values of energy difference.}
 \label{r2dirichlet}
\end{figure}

Now the 2-point correlation function at small $\Omega$ will differ from
Eq.~(\ref{162}). Indeed, formal calculation of the pre-factor (\ref{meana})
shows that it diverges at small $r$ and its leading behavior is the following
\begin{equation}
A\rightarrow \frac{\pi}{2}<r><\frac{1}{\sqrt{r}}>^2.
\end{equation}
But for variable (\ref{rrandom})
\begin{equation}
 <r>=\frac{4}{\pi^2 ab}(\int_0^{\pi} \sin^2\phi\  d\phi)^2=\frac{1}{ab},
\end{equation}
and
\begin{equation}
<\frac{1}{\sqrt{r}}>=
  \frac{\sqrt{ab}}{2\pi^2}(\int_{\phi_0}^{\pi}\frac{d\phi}{\sin \phi})^2
\approx \frac{\sqrt{ab}}{2\pi^2} \ln^2 \phi_0 ,  
\end{equation}
where $\phi_0$ is a cut-off of the integration over $\phi$. With logarithmic
accuracy $\phi_0$ is proportional to 
$\Omega$, $\phi_0\rightarrow \Omega/\Omega_0$,
and when $\Omega \rightarrow 0$
\begin{equation}
r_2(\Omega)\rightarrow \frac{1}{8\pi^3}\Omega \ln^4 (\Omega/\Omega_0).
\label{smallomegad}
\end{equation}
The results of numerical calculations of 100000 levels of the Seba billiard 
with Dirichlet boundary conditions (with irrational ratios (\ref{ratio}) 
and $v'=1$) are presented at Figs.~\ref{nsdirichlet} and \ref{r2dirichlet}. 
In Fig.~\ref{nsdirichlet}
the cumulative nearest-neighbor distribution is plotted. The dashed and
thin solid lines are the same as in Fig.~\ref{nsperiodic}, the dotted line
corresponds to the Poisson distribution
\begin{equation}
N_P(s)=1-e^{-s}.
\end{equation}
Note that the computed distribution is quite far from all standard examples.
Though the resulting distribution is more close to the Poisson distribution
than the one for the Seba billiard with periodic boundary conditions (see
Fig.~\ref{nsperiodic}) one can check that this difference will be present for
all non-zero values of the coupling constant (and, in particular,
when $v'\rightarrow \infty$).

The 2-point correlation function is shown in Fig.~\ref{r2dirichlet}. The
limiting behavior for small (see (\ref{smallomegad})) and large (see
(\ref{asr2d})) values of arguments are also indicated for comparison  by
thick solid lines. The value of $\Omega_0$ in Eq.~(\ref{smallomegad}),
$\Omega_0=52.25$, has been obtained by fitting expression (\ref{smallomegad})
to numerical result for small $\Omega$.

\section{Conclusion}

We have computed analytically the 2-point correlation function for zeros of
random meromorphic functions with large number of poles when these poles are
independent random variables. It is demonstrated that the statistics of these 
zeros corresponds to a distribution with level repulsion which differs from
known examples. The resulting distribution is not universal but depends on
residues. 

A natural example, where energy levels obey such meromorphic equation,
corresponds to a rectangular billiard with a short-range impurity and our
results give the spectral statistics of these models. It is of interest that
different boundary conditions give very different results. Even the
asymptotic behavior for small energy difference is different (cf.
Eqs.~(\ref{smallomega}) and (\ref{smallomegad})).

We also proposed methods which permit to find the behavior of the
2-point correlation function at small and large arguments without the
knowledge of the exact solution. The methods can be applied for cases
where  exact solutions are not available.

\section{Appendix.}

The purpose of this Appendix is to compute the main integral 
(\ref{mainI})
\begin{equation}
  I(\alpha_1,\alpha_2)=\int_{-W}^{W}(1-\exp(i\frac{\alpha_1}{E_1-e}+
  i\frac{\alpha_2}{E_2-e}))de .  
\end{equation}

Let us first derive a few useful relations.  
\begin{equation}
(\frac{\partial}{\partial \alpha_1}-\frac{\partial}{\partial \alpha_2})
J(\alpha_1,\alpha_2)=i\omega \int_{-\infty}^{\infty}
\exp(i\frac{\alpha_1}{E_1-e}+i\frac{\alpha_2}{E_2-e})\frac{de}{(E_1-e)(E_2-e)},
\end{equation}
where
\begin{equation}
\omega= E_1-E_2.
\end{equation}
Perform in this integral the change of variable
\begin{equation}
E_1-e=\frac{\omega}{1+t}.
\end{equation}
Now
\begin{equation}
E_2-e=-\frac{t\omega}{1+t},\;(E_1-e)(E_2-e)=-\frac{t\omega^2}{(1+t)^2},
\;de=\frac{\omega}{(1+t)^2}dt.
\end{equation}
Hence
\begin{equation}
(\frac{\partial}{\partial \alpha_1}-\frac{\partial}{\partial \alpha_2})
J(\alpha_1,\alpha_2)=-i\exp(i\frac{\alpha_1-\alpha_2}{\omega})\int_{-\infty}^{\infty}
\exp(i\frac{\alpha_1}{\omega}(t-\frac{\alpha_2}{\alpha_1 t}))\frac{dt}{t}.
\end{equation}
The following standard integrals will be useful for us (see \cite{2}). When
Im$\mu>0$ and Im$\beta^2\mu<0$
\begin{equation}
\int_0^{\infty}t^{\nu-1}
\exp(i\frac{\mu}{2}(t-\frac{\beta^2}{t}))dt=2\beta^{\nu}
e^{i\nu \pi/2}K_{-\nu}(\beta \mu),
\end{equation}
and when Im$\mu>0$ and Im$\beta^2\mu>0$
\begin{equation}
\int_0^{\infty}t^{\nu-1}\exp (i\frac{\mu}{2}(t+\frac{\beta^2}{t}))dt=
i\pi\beta^{\nu}e^{-i\pi\nu/2}H^{(1)}_{-\nu}(\beta \mu).
\label{56}
\end{equation}
Here $K_{\nu}(x)$ and $H_{\nu}^{(1)}(x)=J_1(x)+iY_1(x)$ are respectively the Macdonald
and Hankel functions. 

Note that $K_0(x)$ is a real function, therefore 
\begin{eqnarray}
   &&\int_{-\infty}^{\infty}
   \exp(i\frac{\alpha_1}{\omega}(t-\frac{\alpha_2}{\alpha_1 t}))\frac{dt}{t}
   =\int_{0}^{\infty}
   \exp(i\frac{\alpha_1}{\omega}(t-\frac{\alpha_2}{\alpha_1
     t}))\frac{dt}{t}-c.c.
   \nonumber\\   
    &&=\left \{\begin{array}{ll}
    0,&\mbox{if}\; \alpha_1\alpha_2>0\\
    2\pi i J_0(-\frac{2}{\omega}\sqrt{\alpha_1\alpha_2})\mbox{sgn}(\alpha_1),
    &\mbox{if}\; \alpha_1\alpha_2<0
\end{array} \right . ,
\end{eqnarray}
and $J_0(x)$ is the Bessel function of zero order.
Finally
\begin{equation}
(\frac{\partial}{\partial \alpha_1}-\frac{\partial}{\partial \alpha_2})
J(\alpha_1,\alpha_2)=
\exp(i\frac{\alpha_1-\alpha_2}{\omega})\Phi(\alpha_1,\alpha_2),
\end{equation}
where
\begin{equation}
\Phi(\alpha_1,\alpha_2)=2\pi  
J_0((\frac{2}{\omega}\sqrt{-\alpha_1\alpha_2})
\Theta(-\alpha_1\alpha_2)\mbox{sgn}(\alpha_1),
\label{59}
\end{equation}
$\Theta (x)=1$ if $x>0$ and $\Theta (x)=0$ if $x<0$.
Note that 
\begin{equation}
\frac{\partial^2}{\partial \alpha_1 \partial \alpha_2} \Phi(\alpha_1,\alpha_2)=
\frac{1}{\omega^2}\Phi(\alpha_1,\alpha_2).
\end{equation}

Using the same method one can prove the following set of equalities
\begin{eqnarray}
  \frac{\partial^2}{\partial \alpha_1 \partial \alpha_2}
  J(\alpha_1,\alpha_2)&=&
  -\frac {i}{\omega}\exp(i\frac{\alpha_1-\alpha_2}{\omega}) \Phi(\alpha_1,\alpha_2).
\\
  \frac{\partial^2}{\partial \alpha_1^2} J(\alpha_1,\alpha_2)&=&
   \exp(i\frac{\alpha_1-\alpha_2}{\omega})\frac {\partial}{\partial \alpha_1} 
   \Phi(\alpha_1,\alpha_2). \label{sd}
\\
  \frac{\partial^2}{\partial \alpha_2^2} J(\alpha_1,\alpha_2)&=&
   -\exp(i\frac{\alpha_1-\alpha_2}{\omega})\frac {\partial}{\partial \alpha_2} 
   \Phi(\alpha_1,\alpha_2).
\\
  \frac{\partial^4}{\partial \alpha_1^2\partial \alpha_2^2}
  J(\alpha_1,\alpha_2)&=&-
  \frac{1}{\omega^2}(\frac{\partial}{\partial \alpha_1}-\frac
  {\partial}{\partial \alpha_2})  
   [\exp(i\frac{\alpha_1-\alpha_2}{\omega})\Phi(\alpha_1,\alpha_2)].
\end{eqnarray}
Note that the  differentiation of  function $\Phi$ produces
the $\delta$-functions coming from the factor
\begin{equation}
  \Theta(-\alpha_1\alpha_2)\mbox{sgn}(\alpha_1)=
  \frac{1}{2}(\mbox{sgn}(\alpha_1)-\mbox{sgn}(\alpha_2)).
\end{equation}
From Eq.~(\ref{sd}) it follows that the second derivative
\begin{equation}
\frac{\partial^2}{\partial \alpha_1^2} J(\alpha_1,\alpha_2)=\phi_1(\alpha_1,\alpha_2)  
\end{equation}
is known. Therefore
\begin{equation}
J(\alpha_1,\alpha_2)=J(0,\alpha_2)+\frac{\partial J(0,\alpha_2)}{\partial
  \alpha_1}+\int_0^{\alpha_1}(\alpha_1-y)\phi_1(y,\alpha_2)dy,
\end{equation}
where
$J(0,\alpha_2)$ and $\partial J(0,\alpha_2)/\partial \alpha_1$ are
values of $J$ and its first derivative at $\alpha_1=0$. 

In Section \ref{meandensity} it was demonstrated that
$$
J(0,\alpha_2)=\pi |\alpha_2|
$$
and symmetrically
$$
J(\alpha_1,0)=\pi |\alpha_1|.
$$
As the second derivatives equal zero when $\alpha_1\alpha_2>0$ the expression of
$J$ in these regions which is continuous when crossing the $\alpha_1$ and
$\alpha_2$ axis is
\begin{equation}
J(\alpha_1,\alpha_2)=\pi (\alpha_1+\alpha_2)\mbox{sgn}(\alpha_2).
\end{equation}
It is clear that the function $J(\alpha_1,\alpha_2)$ is a continuous
function but with discontinuous first derivatives. The values of these
discontinuities follow from the above discontinuity of the function $\Phi$.

Therefore in the region
$$
\alpha_1>0,\; \alpha_2<0,\;  \alpha_1+\alpha_2<0
$$
(which is the continuation of the lower left square $\alpha_1<0$,
$\alpha_2<0$ through the negative $\alpha_2$ axis)
the function $J(\alpha_1,\alpha_2)$ should take the form
\begin{equation}
J(\alpha_1,\alpha_2)= -\pi( \alpha_2 +\alpha_1)+2\pi \alpha_1 
\exp (-i\alpha_2/\omega)
+\int_0^{\alpha_1}(\alpha_1-y)\phi_0(y,\alpha_2)dy,
\end{equation}  
where 
\begin{equation}
\phi_0(\alpha_1,\alpha_2)=
\exp(i\frac{\alpha_1-\alpha_2}{\omega})\frac {\partial}{\partial \alpha_1} 
   \phi(\alpha_1,\alpha_2),
\end{equation}
and
\begin{equation}
\phi(\alpha_1,\alpha_2)=2\pi  
J_0(\frac{2}{\omega}\sqrt{-\alpha_1\alpha_2})
\end{equation}
coincides with the function $\Phi$ but without the discontinuous factor.

After integration by part and certain transformations one obtains that in
all regions  
\begin{eqnarray}
 && J(\alpha_1,\alpha_2)=\pi(\alpha_1+\alpha_2)\mbox{sgn} (\alpha_2) -
  \pi [i(\alpha_1+\alpha_2)
  G(-\frac{\alpha_2}{\omega},\frac{\alpha_1}{\omega})
  \\&&+ (\alpha_2  J_0(\xi)+i\sqrt{-\alpha_1\alpha_2}J_1(\xi))
\exp (i\frac{\alpha_1-\alpha_2}{\omega})]  
  (\mbox{sgn}(\alpha_1)-\mbox{sgn}(\alpha_2)),\nonumber
\end{eqnarray}
where $\xi=\frac{2}{\omega}\sqrt{-\alpha_1\alpha_2}$ and 
\begin{equation}
G(x,y)=e^{iy}\int_x^{\infty}J_0(2\sqrt{yt})e^{it}dt.
\end{equation}
The function $G(x,y)$ obeys the following relations
\begin{eqnarray}
\frac{\partial G(x,y)}{\partial x}&=&-J_0(\xi)e^{ix+iy},\\
\frac{\partial G(x,y)}{\partial y}&=&-i\sqrt{\frac{x}{y}}J_1(\xi)e^{ix+iy},
\end{eqnarray}
where $\xi=2\sqrt{xy}$. To prove the second identity note that
\begin{equation}
\frac{\partial G(x,y)}{\partial y}=e^{iy}\int_x^{\infty}(iJ_0+
J_0'\sqrt{\frac{t}{y}})e^{it}dt.
\end{equation}
But the integrand equals
\begin{equation}
-i\frac{\partial}{\partial t} (\sqrt{t/y}J_0'(2\sqrt{ty})e^{it})),
\end{equation}
and as $J_0'=-J_1$ one gets the above relation.

The functional equation 
\begin{equation}
G(x,y)+G(y,x)=i+iJ_0(\xi)e^{ix+iy}
\label{functional}
\end{equation}
can be proved by comparing the derivatives of both sides and noting that
the integral (\cite{2}, p.50) 
\begin{equation}
\int_0^{\infty}J_0(at)e^{-\gamma t^2}tdt=\frac{1}{(2\gamma)}
\exp (-\frac{a^2}{4\gamma})
\end{equation}
requires that $G(0,y)=i$.

When $y<x$ the above integral can be indefinitely taken by parts 
which leads to expansion
\begin{equation}
G(x,y)=ie^{ix+iy}\sum_{n=0}^{\infty}(-i\eta)^nJ_n(\xi),
\label{ge}
\end{equation}
where $\eta=\sqrt{y/x}$ and $\xi=2\sqrt{xy}$. 

The above collection of formulae permits to find the expansion of $G(x,y)$
for all values of its arguments. 

Another useful representation of our integral is
\begin{eqnarray}\label{jnew}
  J(\alpha_1,\alpha_2)&=&\pi(\alpha_1+\alpha_2)\mbox{sgn} (\alpha_2)
  \\
 &-&\pi \omega [i(y-x)G(x,y)
 +(x\frac{\partial}{\partial x}-y\frac{\partial}{\partial y})
 G(x,y)]  (\mbox{sgn}(x)+\mbox{sgn}(y)),\nonumber
\end{eqnarray}
where ${x=-\alpha_2/\omega, y=\alpha_1/\omega}$.

\pagebreak

\pagebreak

\pagebreak

\pagebreak

\begin{thebibliography}{99}

\bibitem{porter} {\it Statistical Theory of Spectra: Fluctuations}, edited by 
C.E.  Porter (Academic, New Yourk, 1965).
\bibitem{mehta} M.L. Mehta, {\it Random matrices}, 2nd ed. (Academic Press,
  New York, 1965).
\bibitem{bohigas} O. Bohigas in {\it Chaos and Quantum Physics}, Proceedings
  of the les Houges Summer School of Theoretical Physics, 1989, eds.  M.-J.
  Giannoni, A. Voros, and J. Zinn-Justin (Elsevier, New York, 1991).
\bibitem{berrytabor} M.V. Berry and M. Tabor Proc. R.Soc. A {\bf 349}, 101 
(1976); {\it ibid} J. Phys. A: Math. Gen. {\bf 10}, 371 (1977).  
\bibitem{bohigasgiannoni} O. Bohigas, M .J. Giannoni, and C. Schmit, Phys.
  Rev. Lett. {\bf 52}, 1 (1984).
\bibitem{markloff} J. Marklof, Commun. Math. Phys. {\bf 199}, 169 (1998)
\bibitem{bogomolnybohigas} E. Bogomolny, O. Bohigas, and  P. Leboeuf, J.
  Stat. Phys. {\bf 85}, 639 (1996). 
\bibitem{albeverio} S. Albeverio et al. {\it Solvable models in quantum
    mechanics} (Springer-Verlag, 1988).
\bibitem{seba} P. Seba, Phys. Rev. Lett. {\bf 64}, 1855 (1990).  
\bibitem{bohrmottelson} A. Bohr and B.R. Mottelson, {\it Nuclear Structure}
  (W.A. Benjamin, Inc. New York, 1989), Vol. I, 302.
\bibitem{bogomolnyleboeuf} E. Bogomolny, P. Leboeuf, C. Schmit, Phys, Rev.
  Lett. {\bf 85}, 2486 (2000). 
\bibitem{bogomolnygerland} E. Bogomolny, U. Gerland, and C. Schmit, Phys.
  Rev. E {\bf 59}, R1315 (1999).
\bibitem{gerland} E. Bogomolny, U. Gerland, and C. Schmit, Europ. Phys. Journal
 (2000), to be published.  
\bibitem{2} H. Bateman and A. Erd\'elyi, {\it Higher Transcendental Functions}, Vol II,
  (McGraw-Hill Book Company, 1953).
\bibitem{3} G.N. Watson, {\it A treatise on the theory of Bessel function}, (Cambridge
  Univ. Press. 1962).
\end{thebibliography}
\end{document}